\documentclass[12pt,preprint]{aastex}
\newcommand{\etal}{{et~al.~}}

\newcommand{\co}{\mbox{$^{12}$CO}}
\newcommand{\coa}{\mbox{$^{13}$CO}}

\newcommand{\kms}{\mbox{${\rm km~s}^{-1}$}}

\newcommand{\lsun}{\mbox{$L_\odot$~}}

\newcommand{\sigmasqr}{\mbox{${\sigma}^2$}}
\received{}
\lefthead{Heyer \& Brunt}
\righthead{The Universality of Turbulence in Galactic Molecular Clouds} 

\begin{document}

\def\gtabouteq{\,\hbox{\raise 0.5 ex \hbox{$>$}\kern-.77em
                    \lower 0.5 ex \hbox{$\sim$}$\,$}}
\def\ltabouteq{\,\hbox{\raise 0.5 ex \hbox{$<$}\kern-.77em
                     \lower 0.5 ex \hbox{$\sim$}$\,$}}
\def\vlsr{V$_{LSR}$}
\def\kms{km s$^{-1}$}
%%\def\etal{{et al.}}

%\title{The Common Dynamical State of Galactic Molecular Clouds} 
\title{The Universality of Turbulence in Galactic Molecular Clouds} 

\author{Mark H. Heyer\altaffilmark{1}, 
Christopher M. Brunt \altaffilmark{1}
}

\altaffiltext{1}{Department of 
Astronomy, Lederle Research Building,
University of Massachusetts, Amherst, MA 01003, USA;
heyer@astro.umass.edu,brunt@astro.umass.edu}

\begin{abstract}
The universality of interstellar 
turbulence is examined from observed structure functions of 27 giant 
molecular clouds and Monte Carlo modeling.  
We show that the structure functions, ${\delta}v={v_\circ}l^\gamma$, 
derived from 
wide field imaging of \co\ J=1-0 emission from 
individual clouds are described by a narrow range in the scaling exponent, 
$\gamma$, and the scaling coefficient, $v_\circ$. 
The similarity of turbulent structure functions emphasizes the universaility 
of turbulence in the molecular interstellar medium and accounts for 
the cloud-to-cloud size-line width relationship
initially identified by Larson (1981).  
The degree of turbulence universality is 
quantified by Monte Carlo simulations that reproduce the mean 
squared velocity residuals 
of the observed cloud-to-cloud relationship.  
Upper limits to the variation of the 
scaling amplitudes and exponents for molecular clouds are $\approx$10-20\%.  
The 
measured invariance of turbulence for molecular clouds with vastly 
different sizes, environments, and star formation activity suggests 
a common formation mechanism such as converging turbulent flows 
within the diffuse ISM and a limited contribution of energy from 
sources within the cloud 
with respect to large scale driving mechanisms.
\end{abstract}
\parindent=20pt

\section{Introduction}

Turbulent, non-laminar gas flows are a ubiquitous feature within 
all phases of the interstellar medium (ISM).  Therefore,
accurate descriptions of interstellar turbulence are essential to 
meaningful understanding of ISM dynamics and star formation.  
An important statistical description of fluid dynamics is the generalized 
velocity structure function, 
$$ S_p(l)=<|v(r)-v(r+l)|^p> \eqno(1), $$ 
where $l$ is the spatial displacement between two cells within a 3 dimensional 
volume, $p$ is the order of the structure function,
and the averages are taken over the volume of the fluid.  Over a finite 
spatial range, the structure functions may be described as a power law, 
$S_p(l) \sim l^{\zeta(p)}$ but is often re-framed as an equivalent linear 
expression by taking the pth root, $S_p(l)^{1/p}={\delta}v={v_\circ}l^\gamma$,
where $\gamma=\zeta(p)/p$ is the scaling exponent and $v_\circ$ is the scaling 
coefficient.
The structure function provides a concise description of the 
spatial coherence of velocity differences
within a given region.  Such differences
can arise from both systematic motions (rotation, collapse, outflows) and 
random motions due to turbulent gas flows.  Within interstellar clouds, most 
velocity differences are due to turbulence.

One of the most 
cited and influential studies on interstellar turbulence is Larson (1981) that 
identified a power law relationship between 
the global velocity dispersion, ${\Delta}v$ (\kms), and cloud size, $L$ (pc), 
of molecular clouds 
from values taken from 
the literature, ${\Delta}v=CL^{\Gamma}$ where $C=1.1$ and
$\Gamma=0.38$.  
%%Different variables are used to formally distinguish the 
%%structure function of an individual cloud from the cloud-to-cloud size line
%%width relationship.
%Larson (1981) interpreted 
%this relationship as a signature of 
%Kolmogorov 
%turbulence for which $\gamma$=1/3. 
Using a more homogeneous set of cloud data from the 
Massachusetts-Stony Brook Galactic Plane Survey, Solomon \etal (1987) 
found a similar correlation with a comparable scaling coefficient 
($C=1.0$) but a steeper scaling exponent ($\Gamma=0.5$).
These relationships rely on global velocity dispersions that are distinguished
from velocity differences codified in the structure function of equation (1).
As the sample clouds are 
distributed throughout the Galaxy, 
these do not comprise a singular fluid volume.
Therefore, the connection of these cloud-to-cloud size-line width relationships to 
velocity structure functions of individual clouds would seem remote  
unless one assumes that all clouds 
have approximately the same values for $v_\circ$ and $\gamma$. 
Larson (1981) showed examples of a similar 
scaling law {\it within clouds} by using sizes and velocity dispersions derived from 
molecular tracers with different excitation requirements 
(see also Fuller \& Myers 1992).  However, 
the 
dense regions comprise a small fraction of the mass and 
volume of a molecular 
cloud so the measured velocity dispersions within the spatial scales of 
CS or NH$_3$ emissions may not be 
representative of velocity differences over comparable scales but within a lower
density substrate.  Size-line width relationships derived from clump 
identification algorithms for individual clouds 
exhibit a large degree of scatter with a range of values for $\gamma$ 
(Carr 1987; St\"utzki 
\& G\"usten 1990; Falgarone, Puget, 
\& Pereault 1992) or no 
relationship at all (Loren 1989; Simon etal 2001).  
The absence of a consistent correlation between size and line width of 
embedded structures within a given cloud is due to the limited dynamic 
range of sizes that can be identified by such algorithms 
and the superposition of emission 
from disconnected features along the line of sight (Ballesteros-Paredes \&
Mac Low 2002).
Using structure functions of velocity centroid images, 
Meisch \& Bally (1994)
showed that the scaling exponents are similar for a sample of 4 
molecular clouds but did not consider the variation of the scaling coefficient.
Brunt (2003) 
provides the most convincing evidence for 
similar scaling laws within molecular clouds using Principal Component 
Analysis (PCA) as a tool to recover the true structure function for a given cloud 
(Brunt \& Heyer 
2002).  For each \co\ or \coa\  spectroscopic data cube of a molecular cloud, a set of 
${\delta}v$,$l$ points are determined from the eigenvectors 
and eigenimages.  When the PCA measurements for all clouds are 
combined onto a single plot,
these define
a nearly co-linear set of points.  Such a correlation necessarily results from 
narrow distributions of the 
scaling exponent, $\gamma$,  and coefficient, $v_\circ$, 
for this sample of clouds. 

In this {\it Letter}, we extend the study of Brunt (2003) to 
demonstrate that {\it Larson's cloud-to-cloud 
scaling law is explained only if the structure functions for individual 
clouds are nearly identical.} 
PCA-based ${\delta}v,l$ relationships 
are presented 
that demonstrate the same functional form for structure functions 
for molecular clouds that span 
a wide range in size and environmental conditions.  Monte Carlo models
are constructed that place upper limits to the variation of the 
scaling coefficient and exponent.  Finally, we discuss the consequences
of an invariant turbulent spectrum in the context of the formation 
of interstellar molecular clouds, sources of turbulent energy, and star 
formation.

\section{The Composite Structure Function}

Following Brunt \& Heyer (2002), PCA is applied to spectroscopic data cubes
of \co\ J=1-0 emission from molecular clouds that are part of recent 
wide field imaging surveys at the Five College Radio 
Astronomy Observatory (Heyer \etal 1998; Brunt \& Mac Low 2004) or targeted 
studies of individual giant molecular clouds.  
%Despite the large optical depths, 
%\co\ emission provides the best tracer of the large scale dynamics of a molecular
%cloud.  
Heyer \& Schloerb (1997) and Brunt (2003) show there is little difference 
in the ${\delta}v,l$ relationships derived from \co\ emission and 
the lower opacity \coa\ emission.
For each cloud, a power law is fit to the ${\delta}v,l$ points to determine 
the PCA scaling exponent, $\alpha_{PCA}$,  and coefficient, $v_\circ$.  
For the sample of 27 molecular
clouds, the mean and standard deviation for 
the scaling exponent are 0.62 and 0.09 respectively.  
Based on models with little or zero intermittency, 
this  
PCA scaling exponent corresponds to a 
structure function exponent equal to $0.49\pm0.15$ (Brunt \etal 2003).  
The 
mean and standard deviation of the scaling 
coefficient are 0.90 \kms\  and 0.19 \kms.
These rather narrow distributions of $\gamma$ and $v_\circ$ re-emphasize 
the results of Brunt (2003) that there is not much variation in the 
structure function parameters between molecular clouds.
In Figure~1, we overlay the PCA ${\delta}v,l$ points from the sample of clouds.
The composite points 
reveal a near-identical form of the inferred structure functions.  The solid line 
shows the power law bisector fit to 
all points, ${\delta}v=(0.87\pm0.02)l^{0.65\pm0.01}$.  This PCA derived 
exponent corresponds to a structure function scaling exponent of 0.56$\pm$0.02.

The global velocity dispersion of 
each cloud and the cloud size are determined from the scales of the 
first eigenvector and eigenimage respectively.
Basically, the global velocity dispersion, ${\Delta}v$, is the value of the 
velocity structure function measured at the size scale, $L$, of the cloud.
These points, marked as filled circles within Figure~1, 
are equivalent to the global values used by 
Larson (1981) and Solomon \etal (1987) that define the 
cloud-to-cloud size-line width relationship.  
A power law bisector to this subset  
of points is ${\Delta}v=(0.96\pm0.17)L^{0.59\pm0.07}$. 
The similarity of this
cloud-to-cloud 
relationship with that of the composite points is a
consequence of the uniformity of the 
individual structure functions.  Within the quoted errors, it is 
also similar to the cloud-to-cloud 
size-line width relationships -- $\gamma \sim \Gamma$ and $v_\circ \sim C$.  
Therefore, {\it Larson's 
global velocity dispersion versus cloud 
size scaling law follows directly 
from the near identical functional form of velocity structure functions
for all clouds.}
If there were 
significant differences of $\gamma$ or $v_\circ$ between clouds, 
then the cloud-to-cloud size-line width relationship would exhibit much larger
scatter than is measured by Larson (1981) and Solomon \etal (1987).

\section{The Degree of Turbulence Universality}

The cloud-to-cloud size-line width relationships measured by Larson (1981)
and Solomon \etal (1987) and the composite structure functions shown in 
Figure~1 do exhibit some degree of scatter about the fitted lines.  
The scatter is quantified by the mean square of the velocity residuals, ${\sigma}_{obs}^2$,
for each data set
where
$$ {\sigma}_{obs}^2 = { {\Sigma_i^{N} ({\Delta}v_i-CL_i^\Gamma)^2} \over {N} } \;\;\;\; km^2s^{-2} \eqno(2), $$
N is the number of clouds in the sample, C  and $\Gamma$ are 
the parameters derived by fitting a power law to the observed ${\Delta}v,L$ 
points. 
The value for ${\sigma}_{obs}^2$ for the sample of clouds in Larson (1981) using 
only the \co\ and \coa\ measurements is 1.41 $km^2 s^{-2}$.  
The Solomon \etal (1987) 
sample is a larger, more homogeneous set of clouds and therefore, provides a
more accurate measure of the variance within the cloud-to-cloud size-line 
width relationship.  The corresponding
${\sigma}_{obs}^2$ is 0.88 $km^2 s^{-2}$.   The value of 
${\sigma}_{obs}^2$ for the ${\Delta}v,L$ points in Figure~1 is 1.93 $km^2 s^{-2}$
and 0.35 $km^2 s^{-2}$ for the composite collection of ${\delta}v,l$ points.

The measured scatter, described by ${\sigma}_{obs}^2$, 
of the size-line width relationships is a 
critical constraint to the degree of invariance of turbulence within 
the molecular interstellar medium. 
The scatter arises from several sources.  There are basic measurement errors
in the global velocity dispersion owing to the 
velocity resolution of the measurements and the cumulative statistical error 
of the individual spectra.  Deriving cloud sizes from complex projected 
distributions 
of the molecular gas may also introduce some scatter.  These measurement
errors are rarely shown in any cloud size-line width plots.
A secondary source of scatter is limited or biased
mapping of the molecular cloud.  If a given  map was limited in angular extent 
and centered on a region within the cloud that is actively forming stars
%\footnote{Such biased mapping was typical for 
%many of the early maps of molecular clouds}
then 
the measured "global" velocity dispersion 
may be broader due to localized, expanding motions 
from HII regions or protostellar outflows.  Such enhanced velocity 
dispersions from biased regions may not 
represent the global velocity dispersion of the entire giant molecular 
cloud, that is typically quite extended with respect to star formation sites
within the cloud.  
An additional source of the observed scatter is the uncertainty of the 
distance to each cloud in the sample.   Distances to molecular clouds,
and correspondingly, cloud sizes, 
are generally not known to
precisions better than 25\%.
Finally, and most important for the subject of this study,
true differences in the scaling coefficient and exponent
would also contribute to 
the observed scatter in the size-line width relationship.  It is this 
component that we wish to constrain as it defines the degree to which 
turbulence is invariant in the ISM.

To gauge the universality of turbulence within the molecular interstellar 
medium, a simple, Monte Carlo model is constructed to place limits on the 
variance of the scaling coefficients and scaling exponents of 
individual clouds.  The structure function for each model cloud is defined by 
the parameters, $v_\circ=<v_\circ>+\epsilon_v$ and 
$\gamma=<\gamma>+\epsilon_\gamma$
where $\epsilon_{v}$ and $\epsilon_{\gamma}$ are drawn from gaussian 
probability distributions with 
standard deviations of $\sigma_v$ and $\sigma_\gamma$ respectively.
Both $\sigma_v$ and $\sigma_\gamma$ implicitly contain contributions from 
the measurement errors, biased mapping, and true variations in turbulence.
In the limit of absolute universal turbulence, infinite precision, 
and 
unbiased imaging, $\sigma_v$ = $\sigma_\gamma$=0.
A size, $L_\circ$, is assigned to each cloud from a 
uniform probability distribution
such that 
$0 < log(L_{\circ}) < 2$, corresponding to cloud sizes between 1 and 100 pc.
A global velocity dispersion, ${\Delta}v$, is determined by evaluating 
the structure function at the assigned size of the cloud,
$$ {\Delta}v = (<v_\circ>+\epsilon_v)L_\circ^{<\gamma>+\epsilon_{\gamma}} \eqno(3) $$
Once ${\Delta}v$ is determined from equation (3),  
a random component, $\epsilon_{L}$, 
is added to the cloud size, $L=L_{\circ}(1+\epsilon_{L})$ 
to replicate an uncertainty in cloud distances where 
$\epsilon_{L}$ is drawn from a uniform probability distribution within the 
percentage range, ${\pm}\sigma_D/D$.
Simulations are run for ${\sigma}_D/D$=0, 0.1, 0.25, and 0.5.
The Solomon \etal (1987) data set
is used as the primary 
observational constraint so 
N=272.  Following the results in Figure~1, 
%\footnote{The original sample of Solomon contains 273 clouds 
%but one entry in their
%table contains a size with 0.00 degree so we exclude 
%this cloud from our 
%analysis.} 
$<v_\circ>$=0.9 \kms, and $<\gamma>$=0.5.  For each set of ${\sigma}_v$ 
and $\sigma_\gamma$ parameters, \sigmasqr\ is calculated,
$${\sigma}^2(\sigma_\gamma,\sigma_v) = { {\Sigma_i^N 
({\Delta}v_i - C_1L_i^{\Gamma_1})^2} \over {N} } \;\;\;\; km^2s^{-2} \eqno(4), $$
where $C_1$ and $\Gamma_1$ are determined from a bisector fit to
the ${\Delta}v,L$ points for a single realization. 
To reduce the statistical errors of the simulation, we calculate the mean value 
of ${\sigma}^2(\sigma_\gamma,\sigma_v)$ from 500 realizations.

The Monte Carlo results are shown in Figure~2 where 
$\sigma_v$ and $\sigma_\gamma$ are normalized by  $<v_\circ>$ and $<\gamma>$ 
respectively to 
reflect the fractional variation.
The heavy solid line shows the value of ${\sigma}_{obs}^2$=0.88 $km^2s^{-2}$
from the Solomon \etal (1987)
sample.
This contour 
defines the locus of $\sigma_v/<v_\circ>$, $\sigma_\gamma/<\gamma>$ 
points that reproduce the observed scatter in the Solomon \etal (1987) 
size-line width relationship.  In the unlikely extreme 
limit that $\sigma_\gamma$=0,
then the most $\sigma_v$ can vary is 18-23\% about the observed value of 0.9
\kms.  Conversely, if $\sigma_v$=0, then $\sigma_\gamma$ can vary 
by less than 9-12\% about its fiducial value of 0.5.  
More realistically, $\sigma_v \neq 0$ and $\sigma_\gamma \neq 0$, 
so the percent variations for both parameters are $\sim$8-12\%. 
We emphasize that these are {\it upper limits} to the true 
variations between clouds
 as 
$\sigma_v$ and  $\sigma_\gamma$ 
also include measurement errors and 
biased mapping effects that are likely present in all cloud-to-cloud size-line 
width relationships.

\section{Implications of Invariant Interstellar Turbulence}

The upper limits to variations of the structure function parameters 
are quite small when one considers the large range 
of molecular cloud environments.  With few exceptions,
giant molecular 
clouds are sites of massive star formation.  The massive stars can, 
in principle, affect
gas dynamics over the extent of a cloud by enhancing the UV radiation field,
driving HII region 
expansions and stellar winds, and are the progenitors of 
supernova explosions.  The star formation activity in smaller clouds is 
generally limited to the birth of low mass stars 
whose impact on cloud dynamics by protostellar winds 
is highly localized and small with respect to the integrated energy input 
from massive stars.
For the sample of molecular clouds used in this study, the 
far infrared {\it IRAS} point source luminosities range from 20 \lsun\ 
(B18, Heiles' Cloud 2, L1228) to 
7.6$\times$10$^5$ \lsun\ (NGC 7538).
Despite the large differences in the amount 
of internal energy injection from newborn stars, 
the functional forms of the turbulent 
structure functions for all clouds are similar.
Either all clouds coincidentally 
redistribute this internal energy into turbulent, 
random motions described by structure functions with the same slopes and 
amplitudes via some self-regulatory process 
or these 
internal energy sources are small compared to an external energy reservoir 
that is common to all regions in the Galaxy.  
Brunt \etal (2004) show that most of the 
turbulent energy of a molecular cloud resides on the largest scales
and that the cloud dynamics are more readily accounted by large 
scale driving of turbulence.  The universality of turbulence described 
in this study provides additional evidence for large scale driving 
sources within the molecular interstellar medium.  The narrow range 
of turbulent flow parameters may also reflect the necessary conditions that 
facilitate the development of molecular clouds.  Regions with extreme,
high values
of $v_\circ$ may be highly overpressured with respect to self-gravity and 
expand to larger, more diffuse configurations with less effective 
self-shielding to sustain significant molecular abundances.

Larson (1981) presaged the results from recent numerical simulations 
that molecular clouds are 
short lived ($<$10$^7$ years), transient objects (Ballesteros-Paredes, 
Hartmann, \& Vazquez-Semadeni 1999) that must 
be continually reassembled from the diffuse gas component.  His suggestion 
that molecular clouds result from thermal instabilities within shocks 
of colliding atomic gas streams is 
similar to recent numerical simulations of compressible turbulence that show
the development of high density regions directly from the shocks
(Hunter \etal 1986; Elmegreen 1993; Ballesteros-Paredes, Vazquez-Semadeni, \& Scalo 1999).
Such a common dynamical origin of 
molecular clouds in the Galaxy may account for the measured 
near invariance of turbulence.

The results presented in this study apply to the low density gas substrate
that comprises most of the mass and volume of a molecular cloud.  Turbulent 
properties may indeed differ between clouds 
within the high density, localized, supercritical 
core 
regions.   In distributed,  low mass star forming regions, the non-thermal 
motions within the dense gas are 
subsonic (Benson \& Myers 1989).  Within the massive cores typical 
of clustered star forming 
regions, the observed velocities 
are 
supersonic (Pirogov \etal 2003).
The identification of processes responsible for such divergent, 
dense gas configurations 
remains one of the primary 
challenges to descriptions of star formation 
(Klessen, Heitsch, \& Mac Low 2000; 
Padoan \& Nordlund 2002; Vazguez-Semadeni, 
Ballesteros-Paredes, \& Klessen 2003; Shu \etal 2004;
Mac Low \& Klessen 2004).  
%Given the similarity of structure
%functions between molecular clouds determined in this study, 
%the role of compressible turbulence in generating such different 
%dense core conditions 
%would seem to be limited.

\section{Summary}
\parindent=20pt
We have examined the degree to which interstellar turbulence is universal 
within the molecular gas component of the Galaxy by comparing the 
measured structure functions for 27
giant molecular clouds.  Despite the large differences in cloud environments
and local star formation activity, the structure functions are 
described by very narrow ranges of the scaling exponent and scaling 
coefficient.  The degree of universality is further constrained by 
Monte Carlo simulations that replicate the observed scatter in the 
Larson scaling law that describes the relationship between global 
velocity dispersion and cloud size.  The near invariance of turbulence 
within the molecular interstellar medium suggests a common formation 
mechanism of molecular clouds such as shocks due to 
colliding gas streams of diffuse, 
atomic material as originally suggested by Larson (1981).  
It also implies that the energy contribution from 
internal sources such as stellar winds and expanding HII regions may be small 
with respect to a common, external component. 

\acknowledgments
We acknowledge useful comments from Richard Larson, John Scalo, Bruce
Elmegreen, Paolo Padoan, and Enrique Vazquez-Semadeni.
This work was supported by NSF grant AST 02-28993  to the Five College
Radio Astronomy Observatory.   

{}

\clearpage

\begin{figure}
\plotone {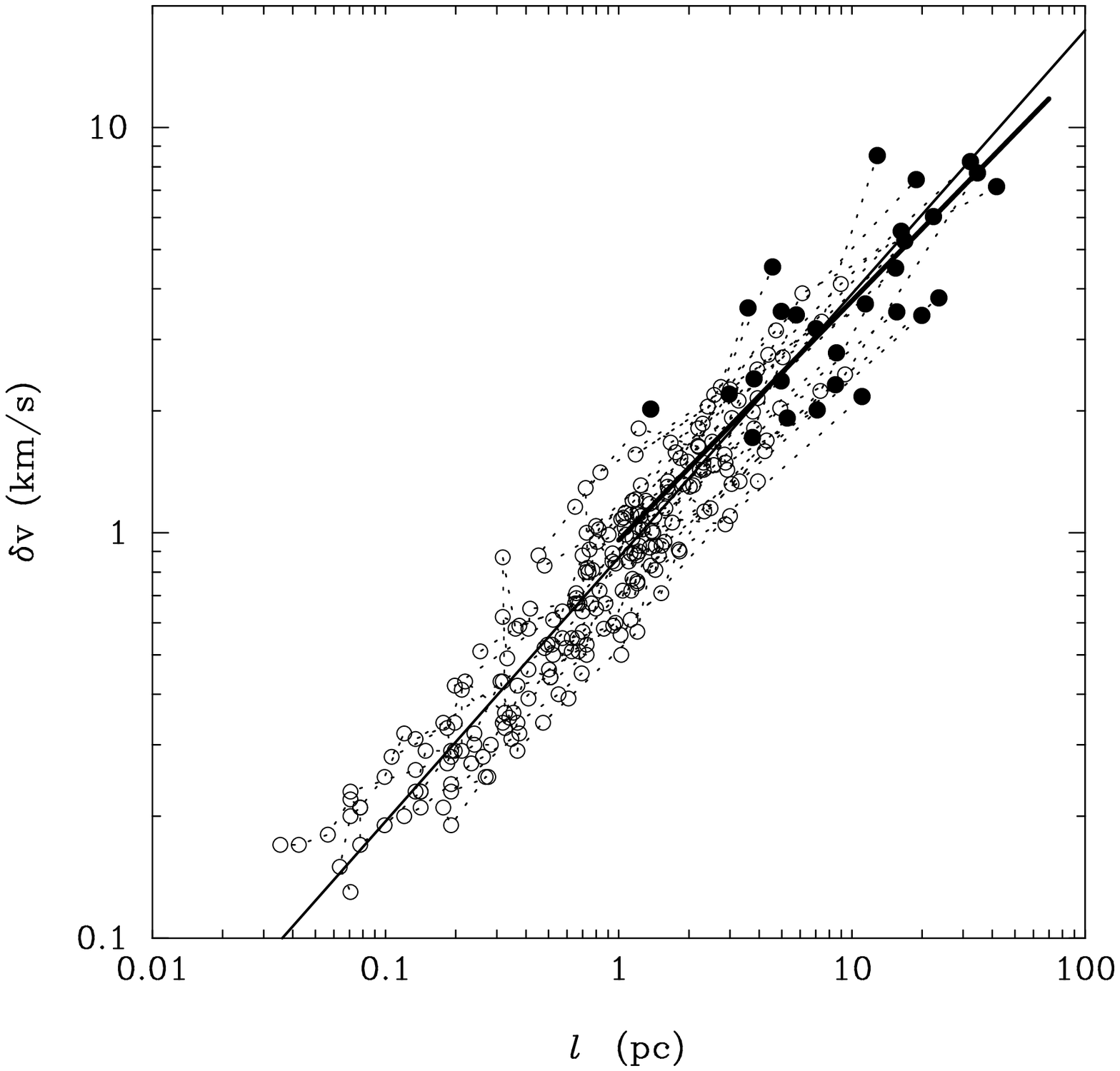}
\caption {The composite ${\delta}v,l$ relationship 
from PCA decompositions of \co\ J=1-0 imaging observations
of 27 individual molecular clouds.
The small scatter of points attest to the 
near invariance of interstellar turbulence within molecular clouds 
that exhibit a large range in size, environment, and star formation activity.
The large filled circles are the global velocity dispersion and size for each 
cloud derived from the first principal component. 
These are equivalent to the global velocity dispersion 
and size of the cloud as would be measured in the cloud-to-cloud size-line 
width relationship (Larson 1981; Solomon \etal 1987).  The light solid line 
show the bisector fit to all points from all clouds.  
The heavy solid line shows the
bisector fit to the filled circles exclusively.   The similarity of these 
two power laws explains the connection of Larson's cloud-to-cloud scaling 
law to the structure functions of individual clouds.
}
\label{composite}
\end{figure}

\begin{figure}
\plotone {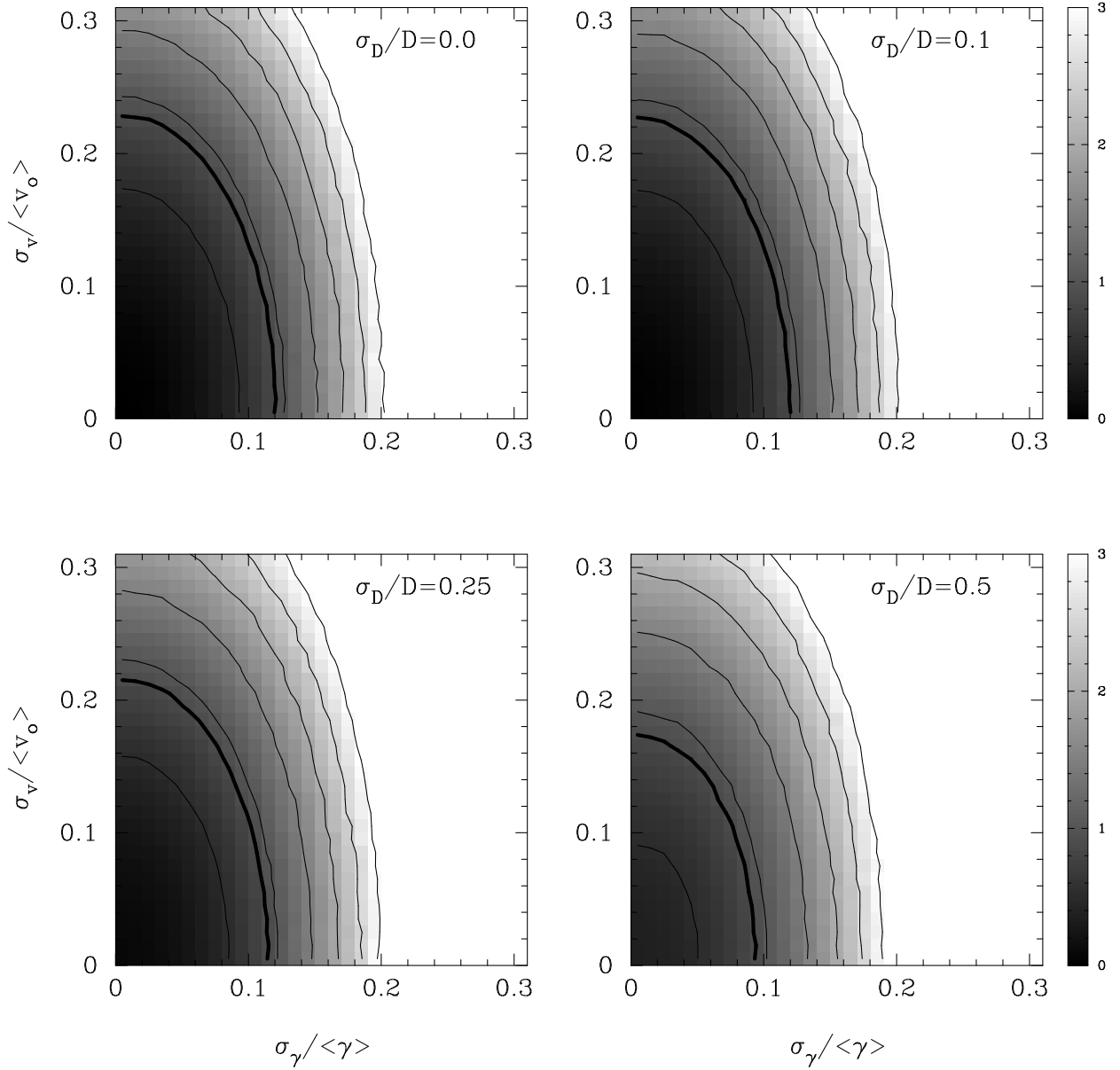}
\caption
{
Images of \sigmasqr\ from the size-line width fits for an 
ensemble of clouds with varying turbulent properties 
parameterized by 
$\sigma_v$/$<v_\circ>$ and $\sigma_\gamma$/$<\gamma>$ for
different distances uncertainties ($\sigma_D/D$=0,0.1,0.25, and 0.5). 
The light contour values of \sigmasqr\ 
are 0.5, 1.0, 1.5 2, 2.5, 3.0 $km^2 s^{-2}$.
The heavy contour in each plot shows the loci that 
correspond to $\sigma_{obs}^2$=0.88 km$^2$ s$^{-2}$ 
from Solomon \etal (1987) and provides
an observational upper limit to cloud-to-cloud variations of 
$v_\circ$ and $\gamma$.  These loci
demonstrate that the structure function parameters, $\gamma$ and $v_\circ$,
can not 
vary by more than $\sim$10-20\%.
}
\label{mc_model}
\end{figure}
\end{document}